\documentclass[aps,superscriptaddress,showpacs,nofootinbib,eqsecnum,prd,notitlepage]{revtex4-1} 

\pdfoutput=1

\usepackage{graphicx}   
\usepackage{subfigure}
\usepackage{hyperref}
\usepackage{float}
\usepackage{color}
\usepackage{amsmath,amsfonts,amsthm} 
\usepackage{slashed}

\newcommand\be{\begin{equation}}
\newcommand\ba{\begin{eqnarray}}
\newcommand\ee{\end{equation}}
\newcommand\ea{\end{eqnarray}}

\begin{document}

\title{Stability of the pion string in a thermal and dense medium}

\author{Arjun Berera} \email{ab@ph.ed.ac.uk} 
\affiliation{School of Physics and Astronomy, 
University of Edinburgh, Edinburgh, EH9 3JZ, United Kingdom}

\author{Robert Brandenberger} \email{rhb@physics.mcgill.ca}
\affiliation{Physics Department, McGill University, Montreal, Quebec, H3A 2T8, Canada, and
Institute for Theoretical Studies, ETH Z\"urich, CH-8092 Z\"urich, Switzerland}

\author{Joel Mabillard} \email{joel.mabillard@ed.ac.uk} 
\affiliation{School of Physics and Astronomy, 
University of Edinburgh, Edinburgh, EH9 3JZ, United Kingdom}

\author{Rudnei O. Ramos} \email{rudnei@uerj.br}
\affiliation{Departamento de F\'{\i}sica Te\'orica, Universidade do
Estado do Rio de Janeiro, 20550-013 Rio de Janeiro, Rio de Janeiro, Brazil}

\begin{abstract} 

We investigate the stability of the pion string in a thermal bath 
and a dense medium. We find that stability is dependent on the
order of the chiral transition. String core
stability within the experimentally
allowed regime is found only if the chiral
transition is second order, and even there the stable region
is small; i.e., the temperature  below which
the core is unstable is close to the critical temperature of the
phase transition.  We also find that the presence of a dense medium,
in addition to the thermal bath, enhances the experimentally accessible 
region with stable strings.
We also argue that once the
string core decays, the ``effective winding'' of the string persists
at large distances from the string core.
Our analysis is done both in the chiral limit, which is
mainly what has been explored in the literature up to now,
and for the physical $h \ne 0$
case, where a conceptual framework is set up for addressing
this regime and some simple estimates are done.

\end{abstract}


\maketitle


\section{Introduction}
\label{sec:intro}

{}From the Grand Unified Theory epoch, where the strong and the
electroweak forces  are expected to have been unified in a single
gauge group, to the later stage of the Standard Model and going
below the energy scale where hadrons are formed, the early Universe is
expected to  have undergone a series of phase transitions. During each
spontaneous breaking of symmetry, it is possible that topological defects are
produced~\cite{vilenkin}. The existence of these defects may explain several open questions in cosmology, such as
primordial density perturbations and structure
formation~\cite{Brandenberger:1993by,Durrer:2001cg}, generation of the
primordial magnetic fields~\cite{Dimopoulos:1997df} and
baryogenesis~\cite{Trodden:1994ve}. Therefore the search 
for topological defects has been an active field of research among 
particle physicists and cosmologists
for the last 30 years.

Embedded defects are a special class of topological 
defects~\cite{Vachaspati:1992pi}. They are constructed by constraining 
a subset of fields in the given theory to vanish, 
while others continue to have solutions of the unconstrained 
system. If the vacuum manifold of the remaining unconstrained part of
the system results in having a
nontrivial homotopy group, then topological defect formation
can occur, with this defect then being 
embedded in the larger theory. Embedded defects are of particular 
interest since they can be constructed in realistic systems in nature. 
Two known examples are the chiral model with the 
pion string~\cite{Zhang:1997is}, which is the focus of this work, and the 
Glashow-Weinberg-Salam model with the 
electroweak string~\cite{Vachaspati:1992fi}. 
The stability of embedded defects is not straightforward and needs
careful analysis, since their existence is not strictly due to the topology of 
the full theory and they are usually not stable in vacuum. The 
field values can escape into the constrained directions and the configuration 
can be continuously deformed to the trivial vacuum. 
Each model needs to be analyzed case by case.

The pion string appears as a special nontrivial solution in chiral models,
in particular the linear sigma model (LSM)
of quantum chromodynamics (QCD).
The pion string corresponds to a classical solution of the LSM
where the charged pions are constrained. Since chiral models are
effective models  commonly used to understand many aspects of QCD, and
in particular used in many investigations related to heavy-ion collision
experiments, one may wonder if pion strings might indeed be
produced in these systems, for example during the quark-gluon plasma to
hadron phase transition in heavy-ion collision
experiments and the early Universe.

The question of the stability of the pion string is of crucial
importance and has been the focus of some previous works.  In
\cite{Nagasawa:1999iv}, Nagasawa and Brandenberger proposed a realistic
mechanism to stabilize the pion string by putting the system in a
thermal bath of photons, whereby interactions of the electromagnetic field
with the charged plasma lead to a lifting of the effective potential
in the constrained field direction. More recent works by Karouby and
Brandenberger \cite{Karouby:2012yz,Karouby:2013vza} confirm
the stabilization effect of this mechanism. Whether this effect is
large enough to have a stable string in the region of parameters that
is experimentally accessible has been the subject of recent discussion
~\cite{Mao:2004ym,Lu:2015yua}. 

In this work, we will study an extension of this stabilization mechanism by placing
the system not only in a thermal environment but also in
a dense medium, which is accounted for by including a nonvanishing
chemical potential.  In addition to the charged plasma, the interactions with fermions
(quarks) will also be included. Thus the model we will
work with is the linear sigma model with quarks (LSMq).
These interactions will then generate further corrections to
the effective potential, making explicit the chiral phase transition
that can occur in the LSMq for instance. These modifications
will lead to a physically more realistic model than has been
studied up to
now~\cite{Mao:2004ym,Lu:2015yua}.    The
string solution will now be altered,
as it depends on the temperature and
the chemical potential.  The analysis 
of the stability to follow will show that the
production of stable strings depends on the order of the chiral phase
transition.

Like topologically stable cosmic strings in local gauge theories,
the pion string is characterized by a core region where the potential
energy is confined. The topology of a string solution can be seen
far beyond this core radius. The stability condition we use in this work 
is that of the stability of the string core against dissipation of the
potential energy from the core region. We will, however, also argue
that even if the string core decays, a remnant of the string solution
persists.

Our analysis finds that the string core stability
condition can be satisfied for experimentally allowed values  when the
chiral transition is second order. In this paper we only consider
classical decay processes. Quantum decay channels have
been studied in~\cite{Karouby:2013vza,Karouby2}.

The paper is organized as follows. In Sec.~\ref{sec:PionStringLinSig}
we briefly review the LSMq at finite temperature and chemical
potential. In the same section we also review the pion string
solution and how it can be stabilized in the context of the
LSMq. Section~\ref{sec:Stability} is dedicated to the stability
analysis of the strings in the thermal and dense medium for
both the chiral limit as well as an initial examination
for the physical case of $h \ne 0$. Our
concluding remarks are given in Sec.~\ref{sec:Discussion}. 


\section{The pion string in the linear sigma model}
\label{sec:PionStringLinSig}

In this section we briefly review the
LSMq~\cite{GellMann:1960np,Caldas:2000ic,Khan:2016exa}
and the pion string solution, which can be seen as an embedded defect
in the LSMq.


\subsection{LSMq at zero temperature}

It is well known that QCD becomes nonperturbative at low energy due to
color confinement. However, the approximate chiral symmetry present in
the QCD Lagrangian and its spontaneous breaking allows one to construct a
low-energy effective theory with hadrons replacing the quarks and
gluons as degrees of freedom. Chiral models, most commonly
the LSMq, have long been used in
many applications aiming to understand various aspects of QCD, among
them, the description of disoriented chiral condensates in heavy-ion
collisions or the chiral phase transition.

The LSMq is a concrete realization of chiral effective theory and
describes interactions between nucleons, pions and sigma fields. We
consider its most simple realization, containing two massless quarks
in a fermionic isodoublet $\psi^\text{T}=(u,\ d)$, a triplet of pseudoscalar pions (${\vec{\pi}}$) and a scalar field sigma ($\sigma$). 
The Lagrangian density of the model reads
\begin{align}
	\mathcal{L}&=\mathcal{L}_\Phi+\mathcal{L}_q,
\label{eq:LagLinSig}
\\  \mathcal{L}_\Phi&=\text{Tr}\left[(\partial_\mu\Phi)^\dagger
(\partial^\mu\Phi)\right]-
m^2\text{Tr}\left[\Phi^\dagger\Phi\right]-\lambda
\left(\text{Tr}\left[\Phi^\dagger\Phi\right]\right)^2+
\frac{1}{2}h\text{Tr}\left[\Phi^\dagger+\Phi\right]\;,
\label{eq:Lagphi}\\ 
\mathcal{L}_q&=\bar{\psi}(i\slashed{\partial}-\gamma^0\mu_q+
g(\sigma+i\vec{\pi}\cdot\vec{\tau}\gamma_5))\psi\;,
\label{eq:Lagq}
\end{align}
where
$\Phi=\sigma\cdot\frac{I}{2}+i\vec{\pi}\cdot\frac{\vec{\tau}}{2}$ is
the meson matrix in Dirac space, $\vec{\tau}$ are the Pauli matrices
with the normalization $\text{Tr}[\tau_a\tau_b]=2\delta_{ab}$ and  $I$
is the identity matrix.  {}Finally, $\mu_q$ is the quark chemical
potential.  The term dependent on $h$ in Eq.~(\ref{eq:Lagphi}) is an
explicit symmetry breaking term.  This term mimics the
breaking of the chiral symmetry in the QCD Lagrangian due to the
nonvanishing quark masses.

In the limit of vanishing $h$, the model has a chiral symmetry
$SU(2)_L\times SU(2)_R$. The spinors
$\psi_{L,R}=\frac{1}{2}(1\pm\gamma_5)\psi$ belong to the fundamental
representation of the group, transforming as
\begin{align}
	\psi_{L,R}\rightarrow
        \exp(-i\vec{\omega}_{L,R}\cdot\vec{\tau})\psi_{L,R}\;.
\end{align}
The scalar fields transform in the $(\frac{1}{2},\frac{1}{2})$
representation,
\begin{align}
	\Phi\rightarrow
        \exp(-i\vec{\omega}_{L}\cdot\vec{\tau})^\dagger \Phi
        \exp(-i\vec{\omega}_{R}\cdot\vec{\tau})\;.
\end{align}
It is easy to check that under such a transformation the Lagrangian
density \eqref{eq:LagLinSig} is invariant. 

The $\Phi$-dependent part of the Lagrangian density is often
explicitly expressed in terms of the pion
($\vec{\pi}\equiv(\pi_0,\pi_1,\pi_2)$) and sigma ($\sigma$) fields,
\begin{align}
	\mathcal{L}_\Phi&=\frac{1}{2}(\partial_\mu\sigma)^2+
        \frac{1}{2}(\partial_\mu\vec{\pi})^2-V_0(\sigma,\vec{\pi})\;,\\  
V_0(\sigma,\vec{\pi})&=\frac{\lambda}{4}(\sigma^2+\vec{\pi}^2-v_0^2)^2-h\sigma\;,
\end{align}	
where $v_0^2=\frac{m^2}{\lambda}\equiv f_\pi^2$ corresponds to the
pion decay constant in the vacuum.

The linear term in \eqref{eq:Lagphi} breaks the chiral symmetry
explicitly by giving a nontrivial vacuum expectation value to the
$\sigma$ field. To construct the classical fundamental state, the
minimum of the potential is considered,
\begin{align}
	\frac{d
          V_0}{d\sigma}&=\lambda(\sigma^2+\vec{\pi}^2-v_0^2)\sigma-h=0\;,\\ 
\frac{d
          V_0}{d\pi_i}&=\lambda(\sigma^2+\vec{\pi}^2-v_0^2)\pi_i=0\;.
\end{align}
The unique solution of the system is
\begin{align}
	\vec{\pi}_0&=0,& & \lambda (\sigma_0^2-v_0^2)\sigma_0=h\;, 
\label{eq:mincond}
\end{align}
and the vacuum expectation value $v$ of the $\sigma$ field to first
order in $h$ reads
\begin{align}
	v&=f_\pi+\frac{h}{2\lambda f_\pi^2}\;.
\end{align}
Assuming that $\sigma=\sigma'+v$, where $\langle\sigma'\rangle_0=0$,
we obtain the shifted Lagrangian density
\begin{align}
	\mathcal{L}_\Phi=&\frac{1}{2}(\partial_\mu\sigma')^2+
        \frac{1}{2}(\partial_\mu\vec{\pi})^2-\frac{1}{2}(-m^2+3\lambda
        v^2)\sigma'^2-\frac{1}{2}(-m^2+\lambda v^2)\vec{\pi}^2-
        \notag\\  &-\lambda\sigma'v(\sigma'^2+\vec{\pi}^2)-
        \frac{\lambda}{4}(\sigma'^2+\vec{\pi}^2)^2-\sigma'(-m^2v+\lambda
        v^3-h)\;,
        \\ \mathcal{L}_q=&\bar{\psi}\left[i\slashed{\partial}-\gamma^0\mu_q+gv+
          g(\sigma'+i\vec{\pi}\cdot\vec{\tau}\gamma_5)\right]\psi\;.
\end{align}	
Note that the term linear in $\sigma'$ vanishes due to \eqref{eq:mincond}. 
In this shifted Lagrangian,
the quarks become massive and the masses of the mesons
are nondegenerate, with vacuum values,
\begin{align}
	m_{q,0}&=g v\;,& m_{\sigma,0}^2&=-m^2+3\lambda v^2\;, &
        m_{\pi,0}^2&=-m^2+\lambda v^2\;.
\end{align}
The parameters $g$, $\lambda$ and $h$ (note that $m^2 = \lambda
f_\pi^2$) are chosen to fit the observable vacuum values,
in particular the pion mass, $m_{\pi,0}=139$ MeV, the pion
decay constant, $f_\pi=93$ MeV, and also the constituent quark
mass $m_{q,0}$ and the mass for the sigma, $m_{\sigma,0}$, whose
values will be explicitly set below.
 
Often the chiral limit of the model is considered. In the absence of the
linear breaking term ($h=0$), the chiral symmetry is spontaneously
broken when the $\sigma$ field develops a vacuum expectation value
$v=v_0\equiv f_\pi$. In the symmetry broken phase the pions become
massless and they correspond to the Goldstone bosons.


\subsection{Chiral phase transition at finite temperature and 
chemical potential}
\label{chiralPT}

The LSMq at finite temperature and chemical potential is expected to
undergo a phase transition in the ($\mu_q$-$T$) plane. {}Following the
arguments of Ref.~\cite{Scavenius:2000qd}, we assume that the most
important contributions to the free energy come from the interactions
with the quarks. The quantum and thermal fluctuations of the meson
fields are neglected (note that this is also a valid assumption in
the large-$N$ approximation for the model~\cite{Andersen:2011ip}).
The (renormalized) free energy or effective potential at one loop
\cite{Caldas:2000ic,Khan:2016exa} reads
\begin{equation}
	V_{\rm eff}(T,\mu_q)=V_0+\Delta V_{0}+\Delta V_{T,\mu_q}\;,
\label{eq:veff}
\end{equation}
where
\begin{eqnarray}
&&V _0=-\frac{1}{2}m^2v^2+\frac{\lambda}{4}v^4-hv, \\ && \Delta V_{0}
  =
  \frac{N_cN_f}{(4\pi)^2}m_q^4\left(\frac{3}{2}+\ln\frac{M^2}{m_q^2}\right)\;,
  \\ && \Delta V_{T,\mu_q} =-2 N_c N_f T \int \frac{d^3k}{(2
    \pi)^3}\left[
    \ln\left(1+e^{-\frac{\omega_k}{T}-\frac{\mu_q}{T}}\right)+
    \ln\left(1+e^{-\frac{\omega_k}{T}+\frac{\mu_q}{T}}\right) \right]\;,
\end{eqnarray}
where $\omega_k=\sqrt{k^2+m_q^2}$, $N_c=3$ is the number of colors,
$N_f=2$ is the number of flavors and $M$ is the regularization scale
used in dimensional regularization in the $\overline{\rm MS}$ scheme.  

The expectation value of the field $\sigma$ in the medium $\langle
\sigma \rangle =v(T,\mu_q)$ corresponds to the minimum of the
effective potential and it is determined by
\begin{align}
	\left.\frac{d V_{\rm eff}}{dv}\right|_{v=v(T,\mu_q)}=0\;,
\label{eq:GapEqu}
\end{align}
which leads to the gap equation,
\begin{align}
	-m^2+\lambda v^2 + \frac{N_c N_f}{4\pi^2}g^4 v^2
        \left[1+\ln\frac{M^2}{g^2v^2}\right]+\frac{N_c N_f}{\pi^2}g^2
        \int_0^\infty dk
        \frac{k^2}{\omega_k}\left[n_F^+(\omega_k)+n_F^-(\omega_k)\right]=
\frac{h}{v}\;,
\label{gapeq1}
\end{align}
where 
\begin{equation}
n_F^\pm=\frac{1}{e^{\frac{\omega_k}{T}\mp\frac{\mu_q}{T}}+1}\;,
\end{equation}
is the Fermi-Dirac distribution for particles and antiparticles.

Let us analyze the chiral limit $h= 0$ and the physical case $h
\neq 0$ separately.


\subsubsection{Chiral limit} 

{}For large $T$ and $\mu_q$ the chiral symmetry is restored.  Equation
(\ref{eq:GapEqu}) is trivially satisfied with  $v(T,\mu_q)=0$, and the
masses of the mesons  are degenerate. The fermions are massless. The
chiral symmetry is spontaneously broken when the  effective potential
develops a nontrivial minimum $v(T,\mu_q) \neq 0$. 

The masses of the mesons $\sigma$ and $\pi$ are given by their
tree-level contributions plus the respective self-energies, which in
our approximation are given by the one-loop corrections due to
the Yukawa interaction,
\begin{align}
	m_{\sigma}^2&=-m^2+3\lambda v^2+\Pi_\sigma^\text{(ren)},
\label{masssigma}
        \\ m_{\pi}^2&=-m^2+\lambda v^2+\Pi_\pi^\text{(ren)},
\label{masspion}
\end{align}
where $\Pi_\sigma^\text{(ren)}$ and $\Pi_\pi^\text{(ren)}$ are the
renormalized one-loop self-energies for the sigma and the pions,
respectively, and given by  (see, e.g., Ref.~\cite{Caldas:2000ic})
\begin{align}
	\Pi_\sigma^\text{(ren)}&=\frac{N_c N_f}{4\pi^2}\left\{g^4v^2
 \left(1+3\ln\frac{M^2}{g^2v^2}\right)+4g^2\int_0^\infty dk        
\frac{k^2}{\omega_k}\left[n_F^+(\omega_k)+n_F^-(\omega_k)\right]
\left(1-\frac{g^2v^2}{\omega_k^2}\right)-\right.\notag\\
	&\left.-4\frac{g^4v^2}{T}\int_0^\infty dk        
\frac{k^2}{\omega^2_k}\left[n_F^+(\omega_k)(1-n_F^+(\omega_k))+
n_F^-(\omega_k)(1-n_F^-(\omega_k))\right]\right\}\;,
\label{eq:selfesig}
\end{align}
and
\begin{align}
	\Pi_\pi^\text{(ren)}&=\frac{N_c N_f}{4\pi^2}\left\{g^4 v^2
        \left(1+\ln\frac{M^2}{g^2v^2}\right)+4g^2\int_0^\infty dk
        \frac{k^2}{\omega_k}\left[n_F^+(\omega_k)+n_F^-(\omega_k)\right]\right\}.
\label{eq:selfepi}
\end{align}

Using Eq.~(\ref{eq:selfepi}) in  the gap equation (\ref{gapeq1}) gives
for $h=0$,
\begin{equation}
	-m^2+\lambda v^2(T,\mu_q)+\Pi_\pi^\text{(ren)} = 0,
\label{Pioncondition}
\end{equation}
which is simply the condition that the pions become massless in the
broken phase, in agreement with the Goldstone theorem.  

We obtain the phase diagram of the model in the ($T,\mu_q$) plane
numerically.  The parameters are fixed by the following conditions:
The vacuum expectation value of the field is $v_0=f_\pi=93 \text{
  MeV}$:
\begin{align}
	\left.\frac{d V_0}{dv}\right|_{v=v_0}=0,
\end{align}
and we require that this minimum is preserved when
quantum corrections are included,
\begin{align}
	\left.\frac{d}{dv}  V_{\rm eff}(T=0,\mu_q=0)\right|_{v=v_0}=0\;.
\end{align}
The mass of the sigma field in the vacuum is in the broad
resonance interval,  $400\,{\rm MeV} \leq m_\sigma \leq 800\,{\rm MeV}$.  
{}For our analysis we set it as
\begin{align}
        m^2_\sigma=\left.\frac{d^2}{dv^2} V_{\rm eff}(T=0,
        \mu_q=0)\right|_{v=v_0}&=(600 \text{ MeV})^2 ,
\end{align}
and for the quark mass we choose
\begin{align}
	m_q&=\left.gv\right|_{v=v_0}=300  \text{ MeV} .
\end{align}
Although there is some freedom in the choice of 
$m_\sigma$ within the
broad resonance interval, this barely influences the stability
of the string.  Thus, we find the following set  of parameters,
\begin{align}
	m^2&=\lambda v_0^2 \simeq (567.7\text{ MeV})^2\;, &
        g&\simeq3.2\;,\notag\\ \lambda&=\frac{1}{2}\left(8\frac{N_c
          N_f}{(4\pi)^2}g^4+\frac{m_\sigma^2}{v_0^2}\right)\simeq37.3\;,&
        M^2&=\frac{m_q^2}{e}\simeq(182.0 \text{ MeV})^2. 
\label{eq:paracl}
\end{align}

\begin{figure}[!h]
	\centering  
\includegraphics[width=8cm]{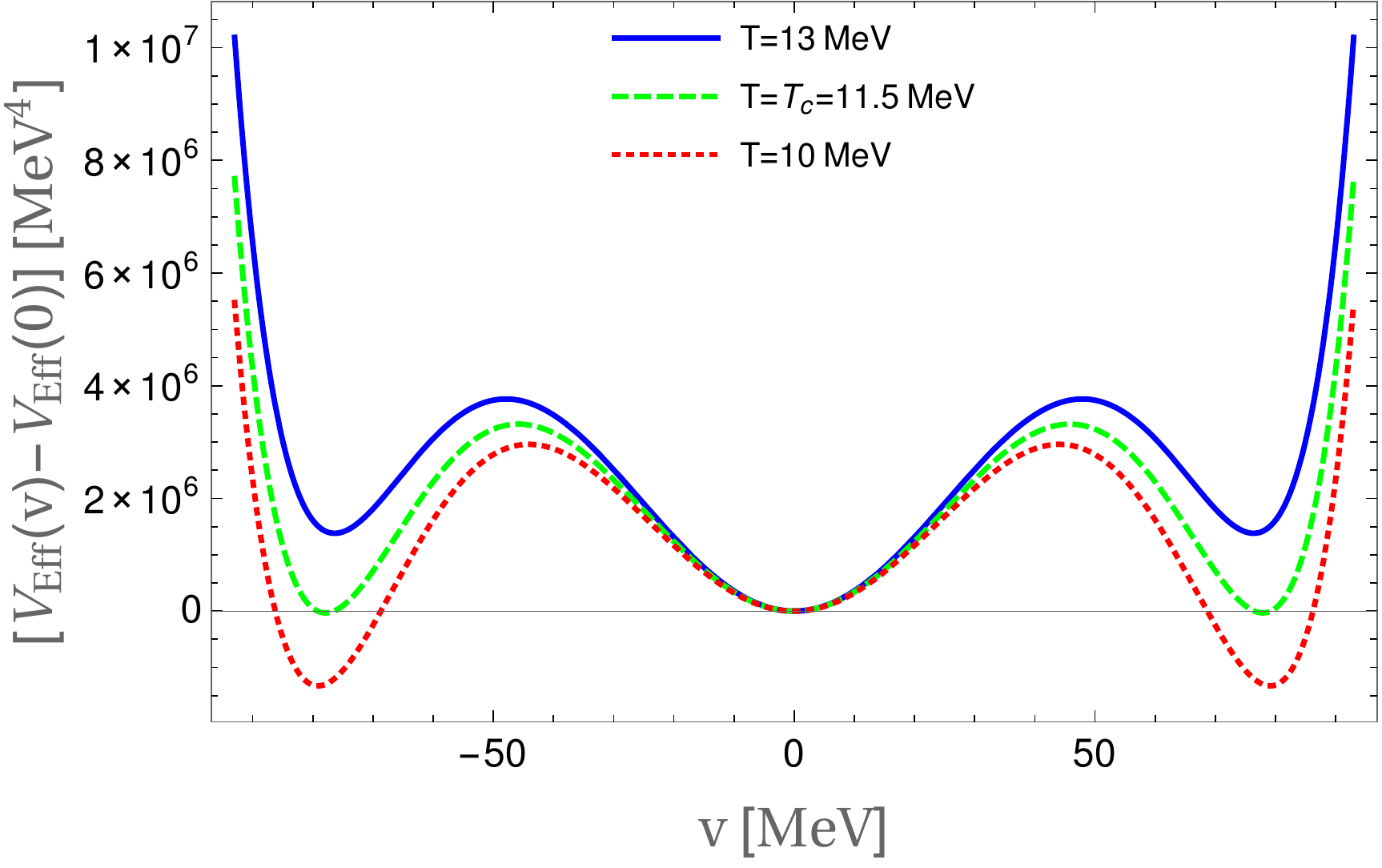}
	  \caption{The effective potential, in the chiral limit, for a
            fixed value of chemical potential $\mu_q=322$ MeV and for
            values of temperature above, at and below the critical
            temperature $T_c$.  Here, $T_c=11.5$ MeV.}
	  \label{fig:FOPT}
\end{figure}

\begin{figure}[!h]
	\centering  
\includegraphics[width=8cm]{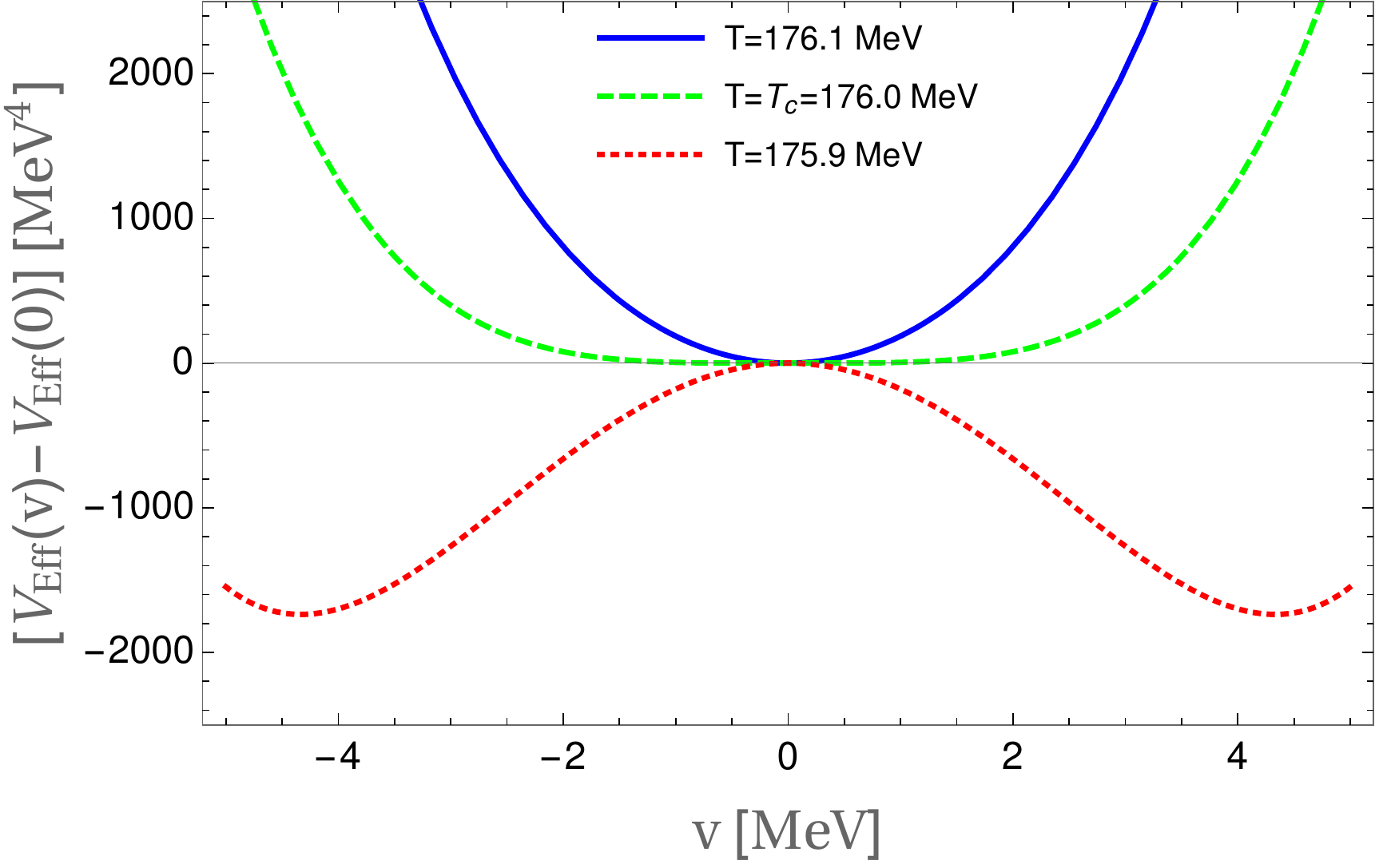}
   \caption{The effective potential, in the chiral limit, for
     $\mu_q=0$ MeV and  for values of temperature  above, at and below
     the critical temperature $T_c$. Here, $T_c=176.0$ MeV.}
   \label{fig:SOPT}
\end{figure}

\begin{figure}[!h]
	\centering 
\includegraphics[width=7cm]{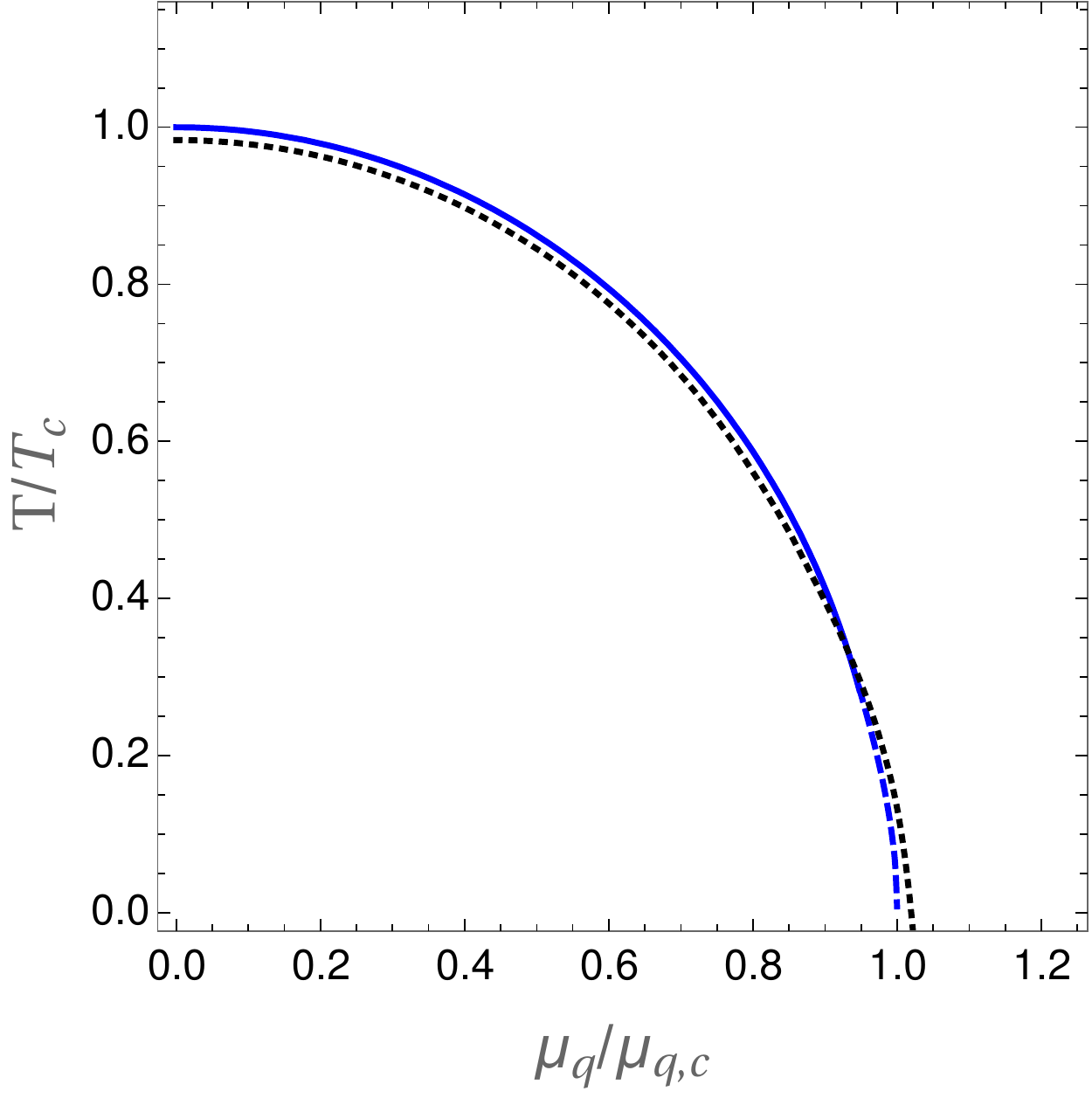}
	  \caption{The phase diagram in the ($\mu_q$-$T$) plane. The
            solid and dashed curves are for the chiral limit ($h=0$)
            and correspond to the second-order and first-order
            transition lines, respectively. The dotted curve is for
            the physical case ($h\neq 0$) and represents a crossover
            transition. Temperature and chemical potential are normalized
by the critical values in the chiral limit: $T_c =176$ MeV and $\mu_{q,c}= 323$ MeV.
{}For the crossover we have the pseudocritical values $T_{pc} = 172$ MeV and 
$\mu_{q,pc}=329$ MeV.}
	  \label{fig:ChiralPT}
\end{figure}

An analysis of the effective potential \eqref{eq:veff} shows that
the order of the phase transition depends on $T$ and $\mu_q$ (which
are related along the phase transition curve). For low
temperatures and large chemical potential, the shape of the effective
potential $V_{\rm eff}$ is typical of a first-order
phase transition, as can be seen in {}Fig.~\ref{fig:FOPT}. In this
case, at $T=T_c$, there are degenerate minima with the origin and the
 expectation value jumps discontinuously at the transition
point.  Then, there is a critical point, which is around
$T=50\ \text{MeV}$ and $\mu_q=306\ \text{MeV}$, above which (as the
temperature increases and the chemical potential decreases) the
phase transition becomes second order. From
{}Fig.~\ref{fig:SOPT} observe that the minimum of the potential moves smoothly
away from zero.  The phase diagram in the ($\mu_q-T$) plane is shown in
{}Fig.~\ref{fig:ChiralPT}. 

\begin{figure}[!h]
	\centering 
\includegraphics[width=8cm]{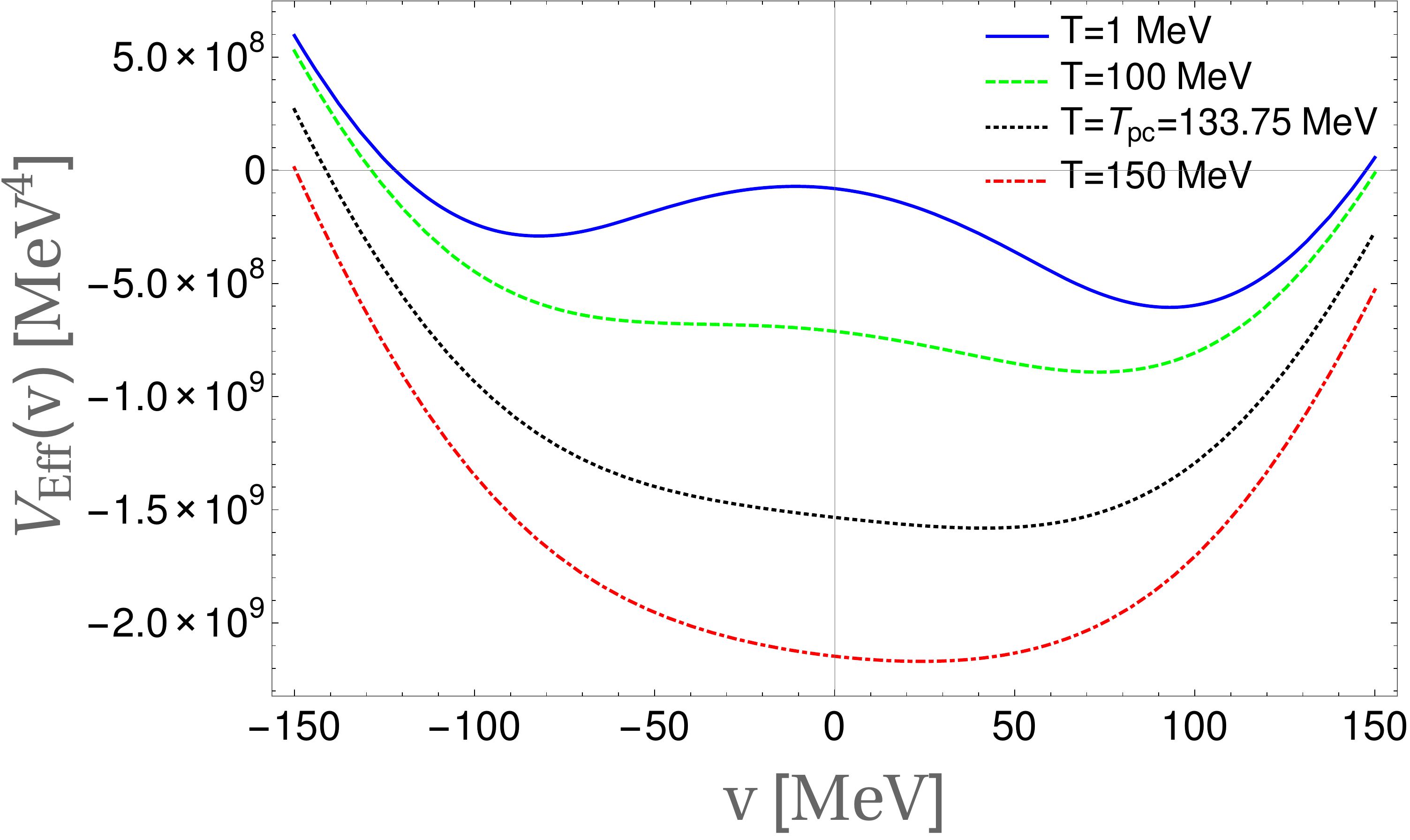}
	  \caption{ The effective potential in the physical case for
            a fixed value of chemical potential $\mu_q=220$ MeV.  It
            shows a crossover phase transition as the temperature is
            changed. There is a pseudocritical temperature at
            $T_{pc}=133.75$ MeV determined by the position of the
inflection point of the $\sigma$ field expectation value.}
	  \label{fig:COPT}
\end{figure}


\subsubsection{Physical case} 

When $h\neq 0$, the symmetry is never completely restored, with
$v(T,\mu_q)$ approaching zero for large values of $T$ and $\mu_q$.
This behavior corresponds to a crossover transition. The gap
equation gives 
\begin{align}
	-m^2+\lambda
        v^2(T,\mu_q)&=-\Pi_\pi^\text{(ren)}+\frac{h}{v(T,\mu_q)} .
\end{align}
The pions are pseudo-Nambu-Goldstone bosons with mass squared
$m_{\pi}^2=\frac{h}{v(T,\mu_q)}$.  The parameters are fixed by the
same requirements as in the chiral limit and the extra condition on
the pion masses in vacuum being set to their physical value
$m_{\pi,0}=139 \text{ MeV}$. {}For this case,
we find the following set  of parameters,
\begin{align}
	m^2&=\lambda v_0^2 -\frac{h}{v_0}\simeq (541.6\text{ MeV})^2\;, &
        g&\simeq3.2\;,\notag\\ \lambda&=\frac{1}{2}\left(8\frac{N_c
          N_f}{(4\pi)^2}g^4+\frac{m_\sigma^2}{v_0^2}-
\frac{h}{v_0}\right)\simeq 36.2\;,&
        M^2&=\frac{m_q^2}{e}\simeq (182.0 \text{
          MeV})^2\;,\notag\\ h&\simeq 1.8\cdot10^6
        (\text{MeV})^3.
\label{eq:parapl}
\end{align}
In the physical case, the effective potential exhibits a crossover
transition, as shown in {}Fig.~\ref{fig:COPT}.
Observe that the minimum of the
potential moves smoothly towards zero as the  temperature
increases. The derivation of the  crossover transition line on the
($\mu_q-T$) plane is performed numerically, with the result depicted in
{}Fig.~\ref{fig:ChiralPT} together with the case for the chiral limit
for comparison.
In our computation, where we have considered both vacuum and thermal
fluctuations for the fermions in the effective potential, we find only a
crossover line. There are though other
approximations where the crossover line can end and merge with a 
first-order phase transition line in a critical end point 
(see, e.g., Refs.~\cite{Khan:2016exa,Scavenius:2000qd}).


\subsection{The pion string solution and its stability}

In Ref.~\cite{Zhang:1997is}, Zhang {\it{et al.}} derived a
stringlike classical solution in the LSM in the chiral limit and in
the vacuum. Defining the new fields $\phi$ and $\pi^\pm$ as
\begin{align}
	\phi&=\frac{\sigma+i\pi^0}{\sqrt{2}},& \pi^\pm&=\frac{\pi^1\pm
          i\pi^2}{\sqrt{2}},
\end{align}
the  $\Phi$-dependent part of the Lagrangian density is rewritten as
\begin{align}
	\mathcal{L}_\Phi&=(\partial_\mu\phi)^*(\partial^\mu\phi)+
(\partial_\mu\pi^+)(\partial^\mu\pi^-)-
        \lambda\left(\phi^*\phi+\pi^+\pi^--\frac{v_0^2}{2}\right)^2.
\label{eq:Lag}
\end{align}
Considering a static configuration, the energy functional, in the
vacuum, reads
\begin{align}
	E_0&=\int
        d^3x\left[\vec{\nabla}\phi^*\vec{\nabla}\phi+\vec{\nabla}\pi^+
\vec{\nabla}\pi^-
          +
          \lambda\left(\phi^*\phi+\pi^+\pi^--\frac{v_0^2}{2}\right)^2\right],
\label{energyE}
\end{align}
and the time-independent equations of motion are
\begin{eqnarray}
	\nabla^2\phi &=&
        2\lambda\left(\phi^*\phi+\pi^+\pi^--\frac{v_0^2}{2}\right)\phi\;,
\label{eqphi}
        \\
        \nabla^2\pi^\pm&=&2\lambda\left(\phi^*\phi+\pi^+\pi^--
\frac{v_0^2}{2}\right)\pi^\pm.
\label{eqpi}
\end{eqnarray}
These equations admit the following pion string solution:
\begin{align}
	\phi&=\frac{v_0}{\sqrt{2}}\rho(r)e^{in\theta}\;,&
        \pi^\pm&=0, 
\label{eq:PionStringSol}
\end{align}
where $r$ and $\theta$ are the polar coordinates in the $(x,y)$ plane
and the integer $n$ is the winding number. The string has a linear extension in
the $z$ direction.

The radial function $\rho(r)$ is found by substituting
Eq.~\eqref{eq:PionStringSol} into the equation of motion and using the
boundary condition,
\begin{equation}
\rho(r) = \left\{
\begin{array}{ll}
0, &\;\;\;r\to 0, \\ 1,& \;\;\;r\to \infty,
\end{array}
\right.
\end{equation}
leading to $\rho(r) \simeq (1-e^{-\mu r})$, where
$\mu^2 = 2 \lambda v_0^2$, and where $\mu^{-1}$ corresponds to the width of the
string. 

The above string solution is, however, nontopological. As it stands,
once formed it will decay away. Since the vacuum manifold of this
model is ${\cal M}=S^3$,  there can be no topological
defects~\cite{vilenkin}. In this case, the nontrivial field
configuration can be continuously deformed to the vacuum.  In other
words, under an infinitesimal excitation of the fields $\pi^\pm$,
the string configuration will unwind. To investigate
the stability of the string, infinitesimal perturbations which involve the $\pi^\pm$ fields
are considered. The perturbations induce a variation of the energy, and
if this variation is negative, the string configuration will be
unstable and decay. This is the case for the above pion string
solution~\cite{Zhang:1997is}.  However, if the effective potential in one of the
field directions is lifted, in particular in the direction of the charged fields
$\pi^\pm$, then we are left with an overall $U(1)$ symmetry of the
effective potential in the $(\sigma,\pi^0)$ directions.  This
then allows for a stable (embedded) topological
pion string to form. This is the case studied by Nagasawa and
Brandenberger  in \cite{Nagasawa:1999iv}, where they proposed a
mechanism to stabilize the pion string by putting the system in a
finite temperature plasma comprised of photons. The interaction between the
charged pions and the electromagnetic field increases the effective
potential in the $\pi^\pm$ directions. The potential for the $\pi^\pm$
fields acquires a quadratic term with a thermal mass  contribution due
to the coupling with the photons~\cite{Karouby:2012yz}, $e^2 T^2 \pi^+
\pi^-/2$, which tends to stabilize the string.


\section{Pion string stability in a thermal and dense medium}
\label{sec:Stability}

\subsection{Setup: Chiral limit}

As shown in Ref.~\cite{Nagasawa:1999iv}, the interactions between the
charged pions and the photons increase the effective potential in the
$\pi^\pm$ directions and act to stabilize the string. We follow the
same strategy but in addition to the thermal bath, we also consider
the  effect of the dense medium due to the interactions with the
fermions.  We assume that the fermions are in equilibrium with the
thermal bath of photons, but, similar to
Ref.~\cite{Karouby:2012yz}, the $\sigma$ and $\vec \pi$ fields are
in a nonequilibrium state. 
  
Using standard techniques~\cite{bellac} a nonzero
chemical potential $\mu_q$ is set for the fermions and the thermal bath is
implemented by the electromagnetic couplings between the charged
particles of the model and the photon. In the minimal coupling
prescription, the Lagrangian density becomes
\begin{align}
\label{eq:lagLSM}
	\mathcal{L}&=\mathcal{L}_\Phi+\mathcal{L}_q-\frac{1}{4}
F_{\mu\nu}F^{\mu\nu}\;,
        \\  \mathcal{L}_\Phi&=(\partial_\mu\phi)^*(\partial^\mu\phi)+
(D^+_\mu\pi^+)(D^{-\mu}\pi^-)-
        \lambda\left(\phi^*\phi+\pi^+\pi^--\frac{v_0^2}{2}\right)^2\;,
        \\  \mathcal{L}_q&=\bar{\psi}\left\{i\gamma_\mu\left[\partial^\mu-
ie\left(\begin{array}{cc}
            q_u & 0  \\  0 &
            q_d \end{array}\right)A^\mu\right]-\gamma^0\mu_q+g(\sigma+
i\vec{\pi}\cdot\vec{\tau}\gamma_5)\right\}\psi.
\end{align}
where $D^\pm_\mu=\partial_\mu \pm i eA_\mu$ and $q_u=2 e/3$,
$q_d=-e/3$ are the electric charges for the {\it{u}} quark and {\it{d}} quark,
respectively.

The interactions with the thermal bath give a thermal mass to the
charged particles, modifying the effective potential in the charged
field directions~\cite{Karouby:2012yz},
\begin{align}
	\left.\Delta V_{\rm eff}\right|_{\text{Thermal Bath}}&=
        \frac{e^2 T^2}{4} \pi^+\pi^-.
\label{pionphoton}
\end{align}
Note also that the coupling to the photons gives a thermal
mass~\cite{bellac} $m_{f}^2(T) = q_f^2 T^2/8$ to the quarks as well.
However, this term can be safely neglected with respect to the $gv$ 
term in the symmetry broken phase.  In addition, at finite temperature 
and chemical potential,
according to the gap equation (\ref{eq:GapEqu}), the expectation value
of the $\sigma$ field is no longer equal to $v_0=f_\pi$, but depends
on $T$ and $\mu_q$, $\langle \sigma \rangle =v\equiv v(T,\mu_q)$. 

In the following we will work in the chiral limit, $h=0$.  To discuss
the pion string in the thermal and dense medium, we use a
mean-field approximation, in particular the Hartree method, 
by integrating out  both the fermions and the
electromagnetic gauge field $A_\mu$.
The Hamiltonian field equations for $\sigma$ and
$\pi_i$, $i=0,1,2$ are found to be
\begin{eqnarray}
	\nabla^2\sigma &=& \lambda \left(\sigma^2 + \vec \pi^2
        - v_0^2 \right)\sigma  +  g \langle {\bar\psi} \psi
        \rangle_{\rm (ren)},
\label{eqsigma1}
\\  
\nabla^2\pi_0&=& \lambda \left(\sigma^2 + \vec \pi^2 -
v_0^2 \right)\pi_0  +  g\langle {\bar\psi} i \gamma_5 \tau_0 \psi
\rangle_{\rm (ren)},
\label{eqpi0}
\\  
\nabla^2\pi_{1(2)}&=& \lambda \left(\sigma^2 + \vec \pi^2 -
v_0^2 \right)\pi_{1(2)}  +  g\langle {\bar\psi} i \gamma_5 \tau_{1(2)}
\psi \rangle_{\rm (ren)} +  e^2\langle A_\mu A^\mu\rangle \pi_{1(2)},
\label{eqp12}
\end{eqnarray}
where we have, in the Hartree-like approximation, 
\begin{align}
	\langle A_\mu \rangle&=0, & \langle A_\mu
        A^\mu\rangle&=\frac{T^2}{4},
\label{AmuAmu}
\end{align}
and by taking the trace of the momentum integral of the fermion propagator, 
the scalar and pseudoscalar fermions densities are~\cite{Scavenius:2000qd} 
\begin{eqnarray}
\langle {\bar\psi} \psi \rangle &=& -2 N_c N_f g \sigma \int \frac{d^3
  k}{(2 \pi)^3} \frac{1}{\omega_k}
\left[1-n_F^+(\omega_k)-n_F^-(\omega_k)\right]\;, \\ \langle {\bar\psi}
i \gamma_5 {\vec\tau} \psi  \rangle &=& -2 N_c N_f g {\vec \pi} \int
\frac{d^3 k}{(2 \pi)^3} \frac{1}{\omega_k}
\left[1-n_F^+(\omega_k)-n_F^-(\omega_k)\right].
\label{psipsi}
\end{eqnarray}
Note that these depend explicitly on the $\sigma$ and $\vec{\pi}$ fields~\cite{Csernai:1995zn}. 
After subtracting the ultraviolet divergent term in the
vacuum-dependent terms of the above momentum integrals, we have
that the finite (renormalized) scalar and pseudoscalar fermion densities are,
respectively,
\begin{eqnarray}
\langle {\bar\psi} \psi \rangle_{\rm (ren)} &=& \sigma \, \Pi_\pi^{\rm
  (ren)}/g\;, \\ 
\langle {\bar\psi} i \gamma_5 {\vec\tau} \psi
\rangle_{\rm (ren)}  &=& {\vec \pi} \, \Pi_\pi^{\rm (ren)}/g,
\label{relations}
\end{eqnarray}
where $\Pi_\pi^{\rm (ren)}$ is given by Eq.~(\ref{eq:selfepi}).

Combining the above Eqs. (\ref{eqsigma1})-(\ref{eqp12}) and  expressing them in terms of
$\phi=(\sigma+i\pi_0)/\sqrt{2}$,  $\pi^\pm= (\pi_1\pm i
\pi_2)/\sqrt{2}$ and also using Eq. (\ref{relations}) together with the
massless pion condition in the chiral limit,
Eq.~(\ref{Pioncondition}), gives
 \begin{align}
	\nabla^2\phi&=2\lambda\left[\phi^*\phi+\pi^+\pi^-
          -\frac{v^2(T,\mu_q)}{2}\right]\phi\;,
        \notag\\ \nabla^2\pi^\pm&=2\lambda\left[\phi^*\phi+\pi^+\pi^--
\frac{v^2(T,\mu_q)}{2}
          + \frac{e^2 T^2}{8\lambda}\right]\pi^\pm.
\end{align}
The above equations generalize the pion string equations in the
vacuum, Eqs.~(\ref{eqphi}) and (\ref{eqpi}).  Hence, the pion string
solution Eq.~(\ref{eq:PionStringSol}) for $\phi$ is modified to
\begin{equation}
	\phi=\frac{v(T,\mu_q)}{\sqrt{2}}{\tilde{\rho}}(r)e^{in\theta},
\label{eq:PionStringSolTmu}
\end{equation}
where $v(T,\mu_q)$ is the solution of the gap
equation~\eqref{eq:GapEqu}, and ${\tilde{\rho}}$ has
the same functional form as $\rho$ except that the inverse
width is now given by $v(T, \mu_q)$. The energy \eqref{energyE} is modified
to
\begin{eqnarray}
E_0 \to E_{ \rm eff} &=&  \int
d^3x\left\{\vec{\nabla}\phi^*\vec{\nabla}\phi+
\vec{\nabla}\pi^+\vec{\nabla}\pi^-
+ \lambda\left[\phi^*\phi+\pi^+\pi^--\frac{v(T,\mu_q)^2}{2}\right]^2
+\frac{e^2 T^2}{4} \pi^+\pi^-\right\}.
\label{Eff}
\end{eqnarray}

\subsection{Stability of the string core}

To investigate the stability of the pion string core, we first consider a
variation of the energy $\delta E$ of the string in the presence of
infinitesimal perturbations of only the charged fields $\pi^\pm$,
\begin{align}
\delta E&= E_{ \rm eff}-E_{\pi^\pm=0}=\int d^3x\left\{\vec{\nabla}\pi^+\vec{\nabla}\pi^- + 
\lambda \left[ \frac{e^2 T^2}{4 \lambda} +
 2 \phi^*\phi-v^2(T,\mu_q)+\pi^+\pi^- \right]
\pi^+\pi^-\right\}.
\end{align}
We use the ansatz \eqref{eq:PionStringSolTmu} and  expand the
perturbations in the direction of  $\pi^\pm$ as
\begin{align}
	\pi^\pm&=v(T,\mu_q) \sum_{m=0}^{\infty} \chi_m(r) e^{\pm im
          \theta}.\label{eq:PionSol2}
\end{align}
Using Eq.~(\ref{eq:PionSol2}),  the variation of the energy in
cylindrical coordinates becomes
\begin{align}
		\delta E&=2\pi v^2(T,\mu_q) \int dz \int r\ dr
                \left\{\chi_m'^2(r)+\frac{m^2}{r^2}\chi_m^2(r)+
                \left[\frac{e^2T^2}{4}
                  + \lambda v^2(T,\mu_q)({\tilde{\rho}}^2(r)-1)+\chi_m^2(r)\right]
                \chi_m^2(r)\right\}.
\label{eq:deltaE}
\end{align}

To determine the stability of the string, it is sufficient to know the 
overall sign of \eqref{eq:deltaE}. A negative variation of energy would imply 
that the string configuration is not favored under an infinitesimal perturbation 
and would likely decay. Considering the integrand of the above equation, the first 
two terms $\chi_m'^2$ and $\frac{m^2}{r^2}\chi_m^2$ are exact squares, so 
necessarily positive (in the next subsection we will explicitly analyze the effect 
of keeping these terms in the stability analysis). The only quantity that may give 
an instability is the last term. A sufficient condition of stability is therefore 
derived from the sign of
\begin{align}
	 \left[\frac{e^2T^2}{4}+ \lambda v^2(T,\mu_q)({\tilde{\rho}}^2(r)-1)+\chi_m^2(r)\right].
\end{align}
The radial function $\chi_m$ is unknown; however, appearing as a square, it 
gives a positive contribution and so for obtaining a minimal
condition for stability it can also be neglected. Using  
${\tilde{\rho}}^2(r)-1 \simeq e^{-{\tilde{\mu}} r}(e^{-{\tilde{\mu}}r} - 2)$ 
[where ${\tilde{\mu}}$ is defined as $\mu$ except that $v_0$ is replaced 
by $v(T, \mu_q)$],  the variation of the mass per unit length compared 
to the embedded string is
\begin{align}
	\frac{e^2 T^2}{4}-2\lambda v^2(T,\mu_q)e^{-{\tilde{\mu}}
          r}(1-\frac{1}{2}e^{-{\tilde{\mu}} r})>0,
\end{align}
or, using that $e^{-{\tilde{\mu}} r}(1-\frac{1}{2}e^{-{\tilde{\mu}} r})\leq \frac{1}{2}$
for all $r$, we find
\begin{align}
	\frac{e^2T^2}{4}-\lambda v^2(T,\mu_q)>0.
\label{eq:DMCond}
\end{align}


\begin{figure}[!h]
	\centering 
\includegraphics[width=7cm]{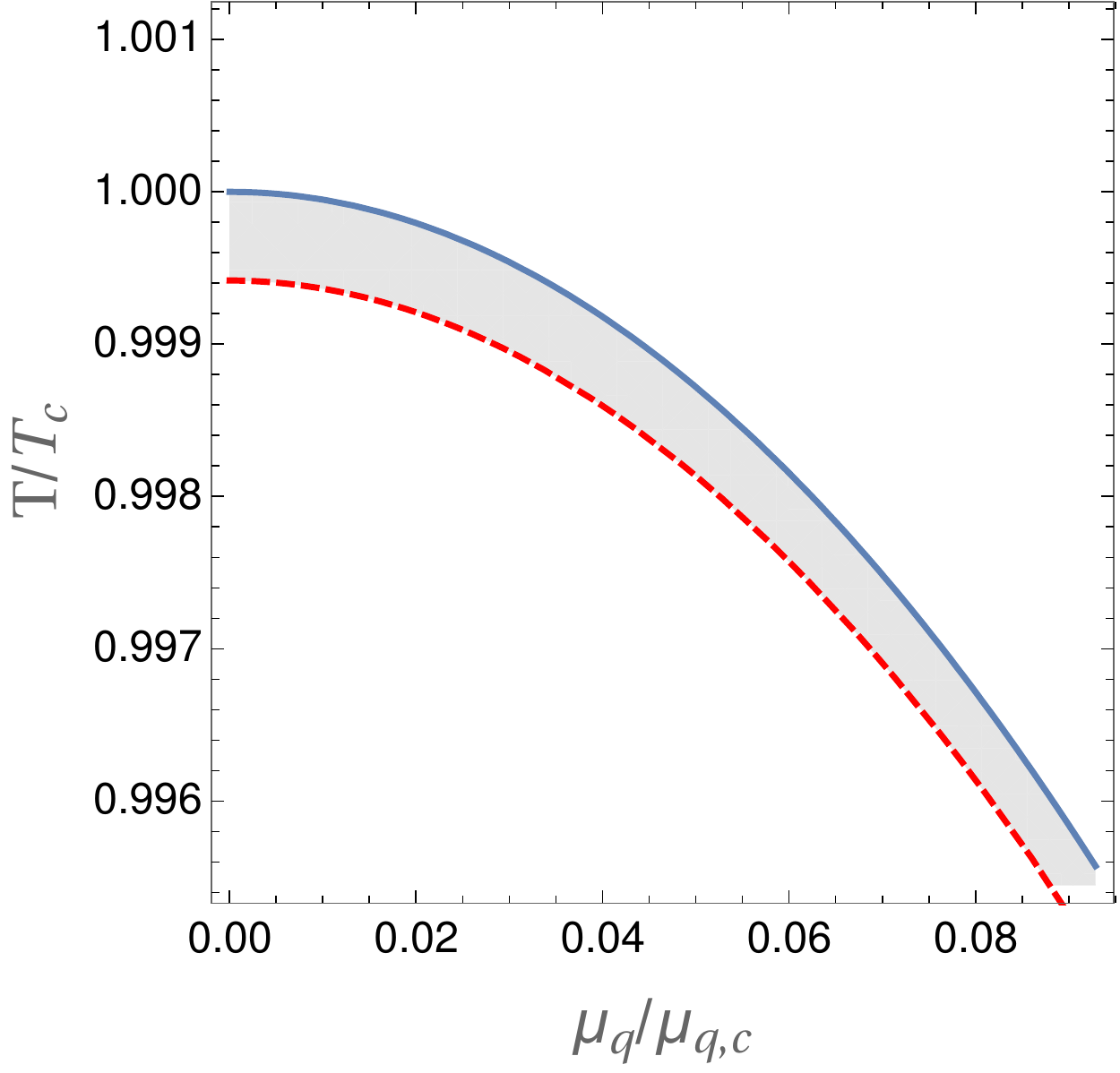}
	  \caption{Stability region of the pion string core in a thermal
            and dense medium in the chiral limit. The parameters are
            those given by Eq.~\eqref{eq:paracl}.  The upper curve (blue)
            corresponds to the phase transition
            (second order). The dashed curve (red) corresponds to the lower
            limit of stability of the string. The range between the
            lines is the region of core stability.}
	  \label{fig:Fcond}
\end{figure}


We compute numerically the region of core stability using the parameters
given in Eq.~\eqref{eq:paracl}. Our results are shown in
{}Fig.~\ref{fig:Fcond}. The top line corresponds to the chiral phase
transition, the string solution being nontrivial in the symmetry
broken phase where $v(T,\mu_q)$ is nonzero.  The dashed line corresponds
to the limit of stability $e^2 T^2/4=\lambda v^2(T,\mu_q)$. The
model predicts a tiny ribbon for values of temperature and chemical
potential, in between the two lines shown {}Fig.~\ref{fig:Fcond}, for
which stable string cores are allowed. 

The size of the stability region is small, but the
following argument makes plausible that such a region
does indeed exist. We know from the results discussed for the LSMq in Sec.~\ref{chiralPT} that the
phase transition is of second order above the critical point.  {}For a
second order phase transition, the expectation value of the field is
exactly zero on the transition line and then it moves away smoothly to
finite values. There is always a region below the phase transition
line where the expectation value $v(T,\mu_q)$ is small enough to
satisfy the stability condition $\eqref{eq:DMCond}$. We, therefore,
expect that the stability condition is always satisfied for a
second-order phase transition. This can be seen explicitly in the
high-temperature approximation.

In the high-temperature region and close to the critical curve, such
that $m_q/T\ll 1$, we use the approximation~\cite{bellac}
\begin{equation}
\int_0^\infty dk \frac{k^2}{\omega_k} \left[ n_F^+(\omega_k) +
  n_F^-(\omega_k) \right] \simeq  \int_0^\infty dk k \left[ n_F^+(k) +
  n_F^-(k) \right]= \frac{\mu_q^2}{2} + \frac{\pi^2 T^2}{6}\;,
\end{equation}
and from the gap equation (\ref{gapeq1}), we find (in the chiral limit
$h=0$ and neglecting the vacuum contribution for simplicity)
\begin{equation}
\lambda v^2(T,\mu_q)\approx \lambda v_0^2 - \frac{N_c N_f}{\pi^2} g^2
\left(\frac{\mu_q^2}{2} + \frac{\pi^2 T^2}{6} \right) \;,
\end{equation}
which using Eq.~(\ref{eq:DMCond})
leads to the approximate analytical stability condition,
\begin{equation}
\frac{e^2T^2}{4} -  \lambda v_0^2 + \frac{N_c N_f}{\pi^2} g^2
\left(\frac{\mu_q^2}{2} +  \frac{\pi^2 T^2}{6} \right) >0 \;.
\label{stab2}
\end{equation}
Values for $T< T_c$ and $\mu_q < \mu_c$ can always be found, i.e.,
temperature and chemical potential below the values corresponding to
those for the critical (second-order) transition line, such as to
satisfy  Eq.~(\ref{stab2}).

The situation changes drastically though when the transition is
first order. It is well known that defects can form during a
first-order phase transition as well (see,
e.g.~\cite{Hindmarsh:1993av}). In our model however, the strings would
decay immediately. The stability condition relies on the smallness of
the temperature and chemical potential background value $v(T,\mu_q)$.
Around the first-order transition the background value $v(T,\mu_q)$
jumps (discontinuously) from zero in the symmetry restored phase to an
usually higher value in the broken phase and the condition
$\eqref{eq:DMCond}$ is never satisfied.  Thus, we conclude that the
existence of stable pion string cores depends strongly on the order of the phase
transition. The stability condition for the pion string is favored
around the second-order transition line of the phase diagram, but it
is disfavored around the first-order transition region. 

The stability condition Eq.~(\ref{stab2}) should be contrasted with
the case where the Yukawa interactions are
absent~\cite{Nagasawa:1999iv}, which leads to
\begin{equation}
\frac{e^2T^2}{4} -  \lambda v_0^2 >0.
\label{stab3}
\end{equation}
Using the values given in Eq.~(\ref{eq:paracl}) and that $e^2 =
4\pi/137$, we find that the temperature of the thermal bath required
for the pion string core stability is
\begin{equation}
T_{\rm stab} > \frac{2 f_\pi \sqrt{\lambda}}{e} \simeq 2.8 \, {\rm
  TeV}.
\end{equation}
This is, however, a temperature way above the critical temperature for
chiral phase transition, $T_c \sim 176$ MeV. Thus, we conclude that it is
simply not possible to have  the stability condition satisfied since
it only happens for temperatures for which the system is already in
the symmetry restored phase. The inclusion of additional thermal and
dense effects from the Yukawa interaction is thus fundamental for
having a stable pion string core.

\subsection{Stability in the physical case $h\neq 0$}
\label{hne0}

In the physical case, $h \ne 0$, 
the effective potential leads to a crossover
transition, as seen in {}Fig.~\ref{fig:COPT}. 
Defect formation in a crossover region is,
unfortunately, very poorly understood at the moment, 
either from analytical studies or from numerical (lattice) simulations.
As far as we know, there is just some limited discussion in
the literature of defect
formation for this case, such as for example Ref.~\cite{Wenzel:2007uh},
where it discusses how defects can be formed by
percolation of different regions with different phases. 

{}For the present case, when accounting only for the background
fields, it would appear that no string solution can be constructed
for the physical case of $h \ne 0$. 
As shown above, e.g. in Eq.~(\ref{eq:PionStringSolTmu}),
the pion string solution is constructed in the plane of the fields 
$(\sigma, \pi_0)$, which is lifted with respect to the charged pions by the
thermal electromagnetic plasma effect. The potential in the plane of the fields 
$(\sigma, \pi_0)$, in the chiral limit $h=0$, is then of the form of a classical
Mexican hat. The string solution interpolates between the unstable vacuum at the
top of the potential to the infinitely degenerate minimum at the bottom of the potential.
The solution then winds around the minima at the bottom of the potential with no cost
of energy. This winding is possible due to the infinitely degenerate minimum of the potential
(the pions are exactly Goldstone bosons).
In the physical case, $h \neq 0$, the chiral symmetry is explicitly broken, the
pions acquire mass and this winding freedom is no longer present
(the potential now becomes a tilted Mexican hat). Under these circumstances,
the string ansatz Eq.~(\ref{eq:PionStringSolTmu}) no longer applies and
for the background fields alone no string solution should be possible to
construct.

The above situation, however, can change significantly when accounting for
fluctuations of the fields in the thermal medium. {}Field fluctuations and gradient 
energies, which are negligible at zero temperature, can grow, particularly close to 
the transition and at large temperatures, where large fluctuations then start to 
become relevant. Under these conditions, it is then feasible that, as these fluctuations 
of the fields grow around the true vacuum of the system (the global minimum of the
potential), they can be sufficiently large to probe the false vacuum state
(the local minimum of the potential). When this happens, we can effectively
say that the winding around the potential is once again restored, at least in localized
regions of space. Much of the system will consist of regions of space
where the fluctuations are small and the state is that of an explicitly chiral
symmetry breaking as usual. However there will some regions with 
larger fluctuations where the chiral symmetry effectively looks restored, and
such regions become increasingly more prevalent
as the temperature increases.  The pion strings that we are interested in are local objects, so all we need is
some suitably large regions where conditions are appropriate for
them to form.  Thus, in regions of large fluctuations, where the chiral
symmetry is effectively restored, pion string formation can become possible once again.
This picture is similar to the mechanism discussed in Ref.~\cite{Wenzel:2007uh} for 
the formation of defects. Typical fluctuations in the fields have a
spatial extent the size of the correlation length,
with $\xi_\sigma^{-1} \sim m_\sigma$ and $\xi_\pi^{-1} \sim m_\pi$. 
As the temperature grows, these fluctuations start to become more and more frequent 
and eventually they start coalescing. In between these regions, string formation is
possible, similar to the Kibble-Zurek mechanism of formation of defects in a 
second-order or even in a first-order phase transition~\cite{Hindmarsh:1993av}.

Though the physics of the formation of these fluctuations in a thermal medium
and their consequences go beyond the analysis allowed within
the framework of the effective
potential,\footnote{ We recall that the computation of the effective potential 
is only able to include the effects of small fluctuations and the proper
treatment requires making use of the effective action
instead. See, e.g., Refs.~\cite{Gleiser:1992ed,Ramos:1996at,Gleiser:2001gy}
for examples of works that try to account for the effect of fluctuations in
a phase transition. Note also that in Ref.~\cite{Mocsy:2004ab} a method has been proposed 
to study the effect of fluctuations in the chiral phase transition in the LSMq,  
without the assumption of the fluctuations to be small.}
we can still provide some reasonable estimates for the
importance of these fluctuation in the present problem.
   
{}Fluctuations in the fields around the true vacuum and that are large 
enough to probe the false vacuum of the potential should have an energy density 
in gradient form comparable to the difference in energy density between the false and true
vacua of the potential, 
\begin{equation}
\langle \frac{1}{2} {\vec \nabla} \sigma. {\vec \nabla} \sigma \rangle +
\langle \frac{1}{2} {\vec \nabla} \pi. {\vec \nabla} \pi \rangle  \approx h v,
\label{grad}
\end{equation}
where we have used that $\Delta V_{\rm eff} \simeq h v$ 
for the energy density difference.
Assuming Gaussian-like (classical) correlation sized fluctuations 
for the fields in the thermal medium, we can then write~\cite{Dziarmaga:1998js}
\begin{eqnarray}
\langle \frac{1}{2} {\vec \nabla} \sigma. {\vec \nabla} \sigma \rangle
&\simeq & \frac{T}{4 \pi^2} \int_0^{m_\sigma} dk \frac{k^4}{k^2 + m_\sigma^2}
\nonumber \\
&=& (3 \pi -8) \frac{m_\sigma^3  T}{48 \pi^2},
\label{gradsigma}
\end{eqnarray}
and analogous for the gradient energy density for the pion field.   
\begin{center}
\begin{figure*}
\subfigure[]{\includegraphics[width=7cm]{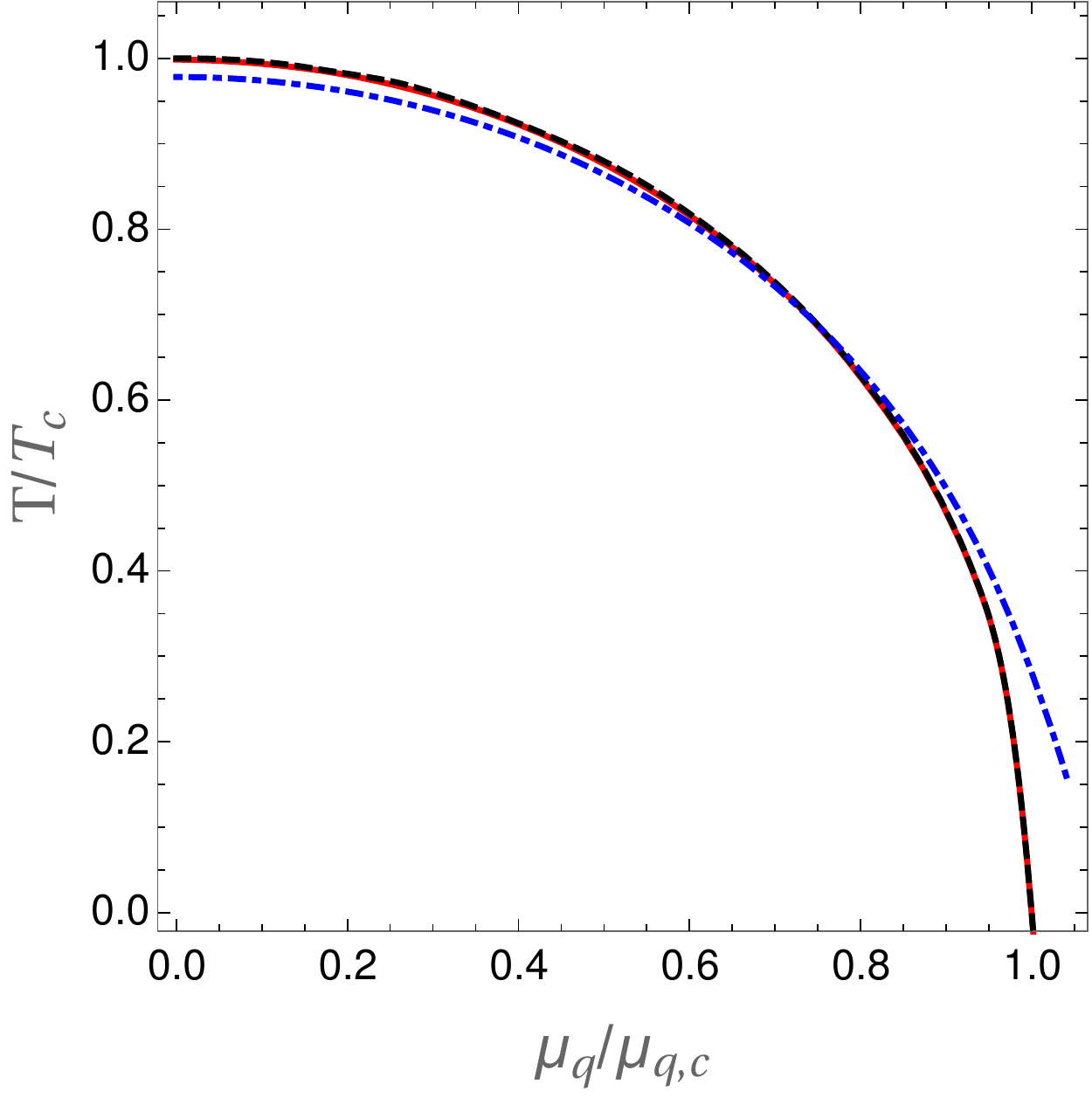}}
\subfigure[]{\includegraphics[width=7cm]{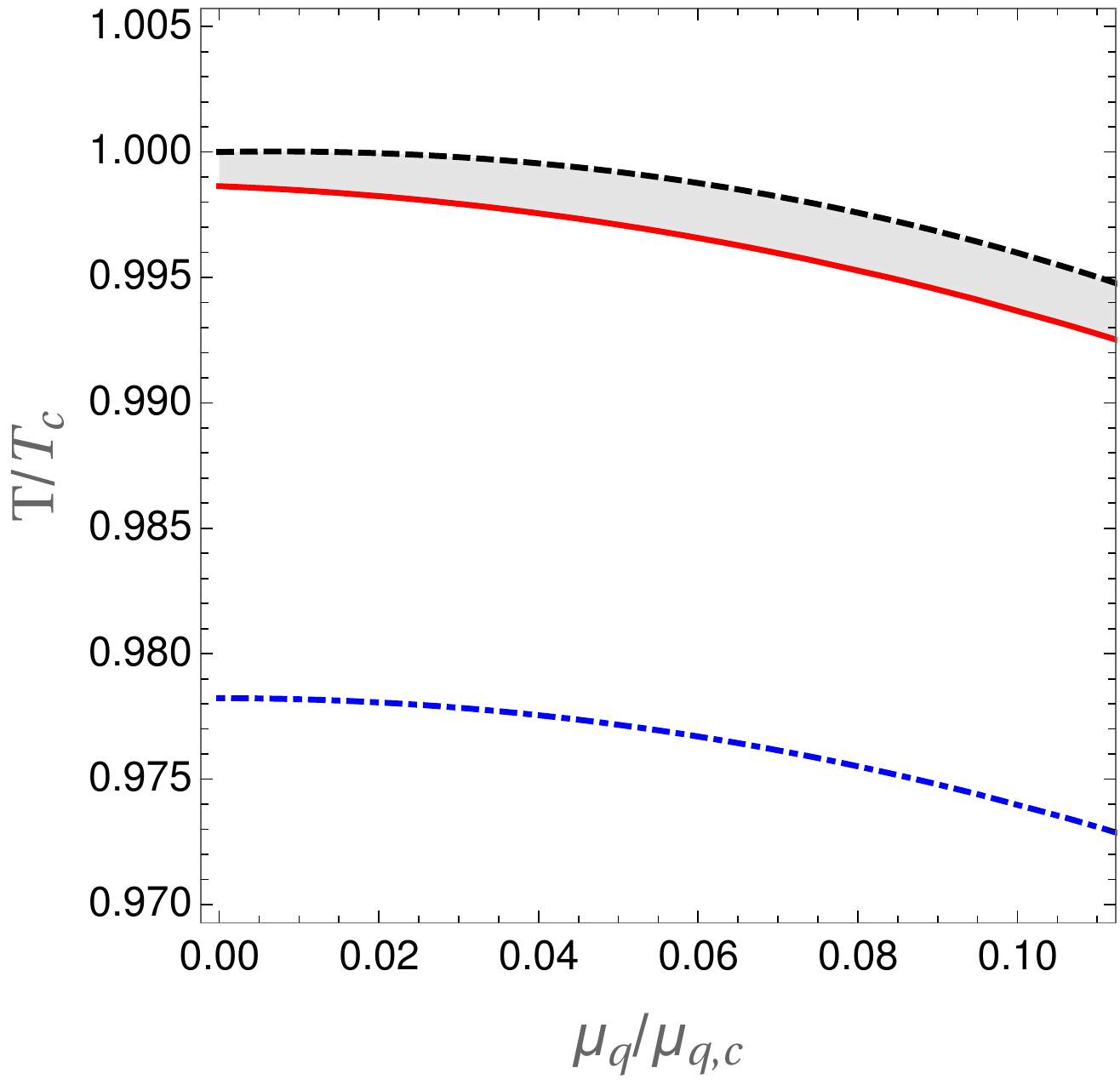}}
\caption{(a) The stability condition for the pion string (red line),
the gradient energy condition (blue dot-dashed line) and the transition line (black dashed line),
in the ($\mu_q,T$) plane (normalized by the corresponding critical values). (b) 
An amplified view around the high-temperature, low chemical potential region.
Strings are allowed to form in the shaded region below the transition line and above
the stability condition. }
\label{fig:Grad}
\end{figure*}
\end{center}
In {}Fig.~\ref{fig:Grad}(a) we show the condition given by Eq.~(\ref{grad})
alongside the transition line and the pion string stability line in the
$(T,\mu_q)$ plane in the physical case of $h\neq 0$. In {}Fig.~\ref{fig:Grad}(b) 
we zoom into a region similar to the one shown previously for the 
chiral limit in {}Fig.~\ref{fig:Fcond}. 
We see from {}Fig.~\ref{fig:Grad}(a) that the gradient energy density
is significantly closer to the transition line and remains slightly below it
down to temperatures and chemical potential around $T\simeq 0.7 T_c$ and
$\mu_q \simeq 0.8 \mu_{q,c}$, when it then goes above
the transition line. 
In this region of large temperatures, the variations in the 
fields are sufficiently large to overcome the difference in potential energy 
density between the local and global minima of the potential.
The stability condition, similar to what we have seen
in the chiral limit $h=0$ (see, e.g., {}Fig.~\ref{fig:Fcond}), is also
very close to the transition line and slightly below it, lying in between
the gradient energy condition and the transition line. 
In this small region
of parameters, in between the stability condition (solid red line) and the transition
line (dashed black line) and lying above the gradient energy condition
(dash-dot blue line), is where pion strings can form
(we locally recover the conditions for winding of the string) and be 
stable at the same time.
Below the line for the gradient energy condition, the fluctuations of the fields 
(in terms of gradient energy) are not large enough to ensure the presence of strings, 
as discussed above. 

The above analysis is just a preliminary examination of the
physical case of $h \ne 0$ and it shows that pion string
formation is plausible in this regime.  More important, this section has laid
out a conceptual framework for how to address this physical regime.
An important general point that this analysis indicates is
the importance that large fluctuations close to the transition
may have on the formation, stability and presence of defects in general.
A complete analysis would require a much more detailed treatment
of the large localized fluctuations that emerge, such
as through numerical simulations,
which is beyond the scope of the present work.
Nevertheless our analysis, though semiquantitative, indicates 
the importance that
gradient energy densities for the fields can have on the pion string
formation and subsequent stabilization when in a thermal and dense medium.
These gradient energy terms can also have important effects in the subsequent 
evolution and decay of these strings when formed, as we discuss below.
Our analysis here also shows that the role of fermions
is an important ingredient to achieve stable pion strings
even in the $h \ne 0$ case due to the effect they
have on the order of the phase transition (recalling that in the absence of
the fermion contributions, no stability is possible
for physically motivated QCD parameters in this model).  Thus the main
focus of our paper on the role of fermions can be seen already
to be important also for any detailed study of the $h \ne 0$ case.


\subsection{Beyond string core stability}

To study the overall stability of the pion string we need to
consider the positive definite terms in $\delta E$ which were neglected
in the previous subsections. They are in particular the radial gradient
energy term of $\chi$ and the contribution of the $(\pi^{+} \pi^{-})^2$
term. The latter in fact blows up as we integrate out to large distances $r$
from the string core, unless $\chi$ goes to zero sufficiently fast for 
large values of $r$. In this case, the radial gradient energy of $\chi$ cannot
be neglected.  Also, if $\chi$ goes to zero at large values of $r$ it
implies that the string winding in the neutral scalar field plane persists.
Thus, even though the string core is unstable, the string will not totally
decay.

To study this issue in more detail we will consider
fluctuations which force the field to remain in the minimum
potential energy density submanifold. Such a fluctuation is
\ba
|\phi|^2 (r) \, &=& \, \rho^2(r) v^2 {\rm cos}^2 \xi(r) \\
\pi^{+} \pi^{-} (r) &=& \rho^2(r) v^2 {\rm sin}^2 \xi(r) \, ,
\ea 
where $v = v(T, \mu_q)$ and the angle $\xi$ labels the
magnitude of the perturbation. We focus on fluctuations which
only depend on the radius since a nontrivial angular dependence
would increase the energy density. For small $\xi$ and for $\xi$ independent
of radius, this fluctuation reduces to the one considered in the
previous subsection.

For this ansatz, the potential energy density vanishes exponentially
for $r > v^{-1}$. However, associated with the nonvanishing value
of $\pi^{+} \pi^{-}$ there is a contribution to the effective potential
which comes from the temperature term. This can be minimized
by having the profile of $\xi(r)$ decay beyond a width which we call $w$.
In this case, the order of magnitude of the thermal effective potential
energy gain $E_{th}$ is
\be
E_{th} \, \sim \, \frac{e^2}{4} T^2 w^3 v^2 \xi^2 \, .
\ee
There is also a radial gradient energy $E_{grad}$ whose order of
magnitude is
\be
E_{grad} \, \sim \, v^2 w \xi^2 
\ee
since the gradient energy density scales as $v w^{-1}$ and the integration
volume as $w^3$. The potential energy from the core region, on the other
hand, decreases as $\xi^2$ increases. If $w > v^{-1}$ the potential energy
$E_{pot}$ has an order of magnitude of
\be
E_{pot} \, \sim \, \lambda v \bigl( 1 - 2 \xi^2 \bigr) \, .
\ee
If $w < v^{-1}$ the change in potential energy is reduced by a factor of
$(w v)^3$. The positive contributions to the total energy are
thus minimized if we set $w \sim v^{-1}$. In this case, the stability
condition of the core region becomes
\be
\frac{\partial}{\partial \xi^2} \bigl( E_{th} + E_{grad} + E_{pot} \bigr) \, > \, 1 \, ,
\ee
which yields
\be \label{limit}
v^2 (2 \lambda - 1) \, < \, \frac{e^2}{4} T^2 \, .
\ee

 If $\lambda \gg  1/2$ (which is the case for our pion string) then this condition
 reduces to the one (\ref{eq:DMCond}) obtained in the previous subsection. However, for
 $\lambda < 1/2$ we find that the string core is stable for all values of the temperature. 
 
 Let us now focus on the case $\lambda > {1/2}$. The above analysis shows
 that the string core will decay out to a radius of at least $w = v^{-1}$ if
 $T < T_d$, where $T_d$ is the temperature when (\ref{limit}) is saturated. But will
 the string decay completely? To answer this question we have to study
 what happens to the total energy change when $w$ increases beyond the
 value $v^{-1}$.  We find that $\delta E$ is negative if
 \be \label{dispwidth}
 w \, < \, \left( \frac{4}{e^2} \right)^{1/3} T^{-2/3} v^{-1/3} \, \equiv \, w_d \, .
 \ee
 Hence, we conclude that the pion string winding remains for distances from
 the core larger than $w_d$. In this sense, the pion string in fact never
 completely decays, but simply undergoes ``core melting.''
 
 In a cosmological context, note that $w_d$ increases less fast as $T$ decreases
 compared to the Hubble radius which scales as $T^{-2}$. A
 pion string scaling solution with mean string separation given by the Hubble
 radius (the scaling solution which describes topologically stable cosmic
 strings) should hence be stable against total annihilation triggered by the
 core decay.

\section{Conclusions}
\label{sec:Discussion}

In this work we have investigated the effect of a thermal and dense
medium on the stability of the pion string.  We have used the LSMq
model to describe the chiral phase transition using realistic
physical parameters.  We have constructed the corresponding pion
string solution  for the model,  which depends explicitly now on the
temperature and the chemical potential. {}Finally, using the
mechanism similar to the one proposed in  Ref.~\cite{Nagasawa:1999iv}, we have analyzed
the stability for the pion strings and have derived a condition for it
to be satisfied. 

We have shown that at low temperatures, the pion string core
will decay via the excitation of charged pion fields. However,
the nontrivial winding of the neutral scalar fields persists at
large distances from the core. In this sense, we should not
speak of the pion string decay, but of pion string core melting.
Whereas for the pion string configuration the energy density
is confined to the core region, after core decay the
energy will mainly be in field gradient energy which is
dispersed out to a width $w_d$ [see (\ref{dispwidth})] which
increases in time as the temperature decreases.

Our results have shown that the existence of a stable string core depends
crucially on the order of the phase transition. Pion strings are
produced  and can become stable when the phase transition is
second order. This happens because the expectation value of the field
in the medium changes smoothly away from zero. In this case the
stability condition is automatically satisfied in a region close to
the transition line. This argument fails when the transition is
first order since now the minimum of the potential can jump
discontinuously to a large value, such that the stability condition no
longer holds.  In this respect the presence of fermions, which 
is a key direction this paper has explored, is crucial.
The inclusion of the fermions indirectly provides
stability, in the sense that fermions  do not 
change the stability condition Eq. (\ref{eq:DMCond})
but they change the order of the phase transition and therefore bring
stability.  This is a key new result of this paper and this is
the first paper to find a stability region for the pion string.
Although most of the analysis was done mainly in the chiral limit,
in Sec.~\ref{hne0} we have done a preliminary analysis also for
the physical case of $h \ne 0$, where we have pointed out how
fluctuations of the fields leading to large gradient energy
densities, can play an important role in the formation and
stability of pion strings in this regime.

The existence of pion strings has direct consequences for cosmology and nuclear physics. 
The region of the ($\mu_q-T$) plane in {}Fig.~\ref{fig:ChiralPT} with a second-order transition  and stable 
strings has large temperatures and a low chemical potential. This region of the plane applies for both the 
early Universe and aspects of heavy-ion collision. The applications of the pion string in the early Universe 
are multiple. One concrete example is the creation of primordial magnetic fields as discussed in 
Ref.~\cite{Brandenberger:1998ew}. Pion strings in heavy-ion collisions experiments have been discussed 
recently in Refs.~\cite{Mao:2004ym,Lu:2015yua}. The production of strings in these kinds of experiments may 
have an influence on the distribution of baryons and one could speculate about their experimental signature.

Another interesting area to investigate would be to find a
further extension of the mechanism that stabilizes the string. In
order to affect the effective potential in the constrained directions,
one needs to act on the charged pions only. One possibility would be to
place the system in an external magnetic field. We leave this as a possible
future work.

Our work has applications beyond the LSMq of the strong interactions.
Similar considerations can be used to study the stability of the 
Z string~\cite{Vachaspati:1992fi},
the embedded string solution made up of the uncharged complex
Higgs field with the charged complex scalar set to zero. An initial study
of the thermal stabilization of the Z string was given in \cite{Nagasawa2}.
Our work shows that the Z string never completely decays, but at most
undergoes core melting.

Looking beyond the Standard Model of strong, weak and electromagnetic
interactions, and to higher temperatures, it would be interesting to study
if there are embedded defects in beyond the Standard Model (BSM) theories which
could be stabilized not only by a photon plasma, but by a plasma of
the gauge fields which are massless above the electroweak symmetry
breaking scale, and above the confinement scale.  BSM theories with embedded
domain wall solutions stabilized by a plasma in the early Universe could
face severe cosmological problems since a single domain wall crossing
our Hubble patch would overclose the Universe.


\acknowledgments  A.B. is supported by STFC.  R.B. would like to
thank the Higgs Centre of the University of Edinburgh for the invitation to
visit, and he wishes to thank the
Institute for Theoretical Studies of the ETH Z\"urich for kind
hospitality during the 2015/2016 academic year. He acknowledges financial support from Dr.  Max
R\"ossler, the Walter Haefner Foundation, the ETH Z\"urich Foundation,
and from a Simons Foundation fellowship.  The research of R.B. is also
supported in part by funds from NSERC and the Canada Research Chair
program.  J.M. is supported by Principal's Career Development Scholarship and Edinburgh 
Global Research Scholarship. R.~O.~R.~is partially supported by  Conselho Nacional de Desenvolvimento Cient\'{\i}fico e
Tecnol\'ogico - CNPq (Grant No. 303377/2013-5) and Funda\c{c}\~ao Carlos Chagas Filho de
Amparo \`a Pesquisa do Estado do Rio de Janeiro - FAPERJ (Grant No. E - 26 / 201.424/2014).



\begin{thebibliography}{99}


\bibitem{vilenkin} A.~Vilenkin and E.~P.~S.~Shellard,  {\it Cosmic Strings and Other Topological Defects},
  (Cambridge University Press, Cambridge, England,  2000).

\bibitem{Brandenberger:1993by} 
  R.~H.~Brandenberger,
  ``Topological defects and structure formation,''
  Int.\ J.\ Mod.\ Phys.\ A {\bf 9}, 2117 (1994)
  doi:10.1142/S0217751X9400090X
  [astro-ph/9310041].

\bibitem{Durrer:2001cg} 
  R.~Durrer, M.~Kunz and A.~Melchiorri,
  ``Cosmic structure formation with topological defects,''
  Phys.\ Rept.\  {\bf 364}, 1 (2002)
  doi:10.1016/S0370-1573(02)00014-5
  [astro-ph/0110348].

\bibitem{Dimopoulos:1997df} 
  K.~Dimopoulos,
  ``Primordial magnetic fields from superconducting cosmic strings,''
  Phys.\ Rev.\ D {\bf 57}, 4629 (1998)
  doi:10.1103/PhysRevD.57.4629
  [hep-ph/9706513].

\bibitem{Trodden:1994ve} 
  M.~Trodden, A.~C.~Davis and R.~H.~Brandenberger,
  ``Particle physics models, topological defects and electroweak baryogenesis,''
  Phys.\ Lett.\ B {\bf 349}, 131 (1995)
  doi:10.1016/0370-2693(95)00214-6
  [hep-ph/9412266].

\bibitem{Vachaspati:1992pi} 
  T.~Vachaspati and M.~Barriola,
  ``A New class of defects,''
  Phys.\ Rev.\ Lett.\  {\bf 69}, 1867 (1992).
  doi:10.1103/PhysRevLett.69.1867

\bibitem{Zhang:1997is} 
  X.~Zhang, T.~Huang and R.~H.~Brandenberger,
  ``Pion and eta strings,''
  Phys.\ Rev.\ D {\bf 58}, 027702 (1998)
  doi:10.1103/PhysRevD.58.027702
  [hep-ph/9711452].

\bibitem{Vachaspati:1992fi} 
  T.~Vachaspati,
  ``Vortex solutions in the Weinberg-Salam model,''
  Phys.\ Rev.\ Lett.\  {\bf 68}, 1977 (1992)
  Erratum: [Phys.\ Rev.\ Lett.\  {\bf 69}, 216 (1992)].
  doi:10.1103/PhysRevLett.68.1977

\bibitem{Nagasawa:1999iv} 
  M.~Nagasawa and R.~H.~Brandenberger,
  ``Stabilization of embedded defects by plasma effects,''
  Phys.\ Lett.\ B {\bf 467}, 205 (1999)
  doi:10.1016/S0370-2693(99)01140-5
  [hep-ph/9904261].

\bibitem{Karouby:2012yz} 
  J.~Karouby and R.~Brandenberger,
  ``Effects of a Thermal Bath of Photons on Embedded String Stability,''
  Phys.\ Rev.\ D {\bf 85}, 107702 (2012)
  doi:10.1103/PhysRevD.85.107702
  [arXiv:1203.0073 [hep-th]].
  
\bibitem{Karouby:2013vza} 
  J.~Karouby,
  ``String melting in a photon bath,''
  JCAP {\bf 1310}, 017 (2013)
  doi:10.1088/1475-7516/2013/10/017
  [arXiv:1212.1723 [hep-th]].
  

\bibitem{Mao:2004ym} 
  H.~Mao, Y.~Li, M.~Nagasawa, X.~m.~Zhang and T.~Huang,
  ``Signal of the pion string at CERN LHC Pb - Pb collisions,''
  Phys.\ Rev.\ C {\bf 71}, 014902 (2005)
  doi:10.1103/PhysRevC.71.014902
  [hep-ph/0404132].

\bibitem{Lu:2015yua} 
  F.~Lu, Q.~Chen and H.~Mao,
  ``Pion String evolving in a thermal bath,''
  Phys.\ Rev.\ D {\bf 92}, 085036 (2015)
  doi:10.1103/PhysRevD.92.085036
  [arXiv:1507.04174 [hep-ph]].

\bibitem{Karouby2} 
  J.~Karouby and A.~M.~Srivastava,
  ``Baryon production from embedded metastable strings,''
  arXiv:1312.0601 [hep-th].
  
\bibitem{GellMann:1960np} 
  M.~Gell-Mann and M.~Levy,
  ``The axial vector current in beta decay,''
  Nuovo Cim.\  {\bf 16}, 705 (1960).
  doi:10.1007/BF02859738
  
\bibitem{Khan:2016exa} 
  R.~Khan, J.~O.~Andersen, L.~T.~Kyllingstad and M.~Khan,
  ``The chiral phase transition and the role of vacuum fluctuations,''
  Int.\ J.\ Mod.\ Phys.\ A {\bf 31}, 1650025 (2016)
  doi:10.1142/S0217751X16500251
  [arXiv:1102.2779 [hep-ph]].


\bibitem{Caldas:2000ic} 
  H.~C.~G.~Caldas, A.~L.~Mota and M.~C.~Nemes,
  ``The Chiral fermion meson model at finite temperature,''
  Phys.\ Rev.\ D {\bf 63}, 056011 (2001)
  doi:10.1103/PhysRevD.63.056011
  [hep-ph/0005180].

\bibitem{Scavenius:2000qd} 
  O.~Scavenius, A.~Mocsy, I.~N.~Mishustin and D.~H.~Rischke,
  ``Chiral phase transition within effective models with constituent quarks,''
  Phys.\ Rev.\ C {\bf 64}, 045202 (2001)
  doi:10.1103/PhysRevC.64.045202
  [nucl-th/0007030].


\bibitem{Andersen:2011ip} 
  J.~O.~Andersen and R.~Khan,
  ``Chiral transition in a magnetic field and at finite baryon density,''
  Phys.\ Rev.\ D {\bf 85}, 065026 (2012)
  doi:10.1103/PhysRevD.85.065026
  [arXiv:1105.1290 [hep-ph]].

\bibitem{bellac} M. Le Bellac, {\it Thermal Field Theory}, (Cambridge
  University Press, Cambridge, England, 1996);\\
J.~I.~Kapusta and C.~Gale,
  {\it Finite-temperature field theory: Principles and applications},
  (Cambridge University Press, Cambridge, England, 2006).
  
  
\bibitem{Csernai:1995zn} 
  L.~P.~Csernai and I.~N.~Mishustin,
  ``Fast hadronization of supercooled quark - gluon plasma,''
  Phys.\ Rev.\ Lett.\  {\bf 74}, 5005 (1995).
  doi:10.1103/PhysRevLett.74.5005
  
\bibitem{Hindmarsh:1993av} 
  M.~Hindmarsh, A.~C.~Davis and R.~H.~Brandenberger,
  ``Formation of topological defects in first order phase transitions,''
  Phys.\ Rev.\ D {\bf 49}, 1944 (1994)
  doi:10.1103/PhysRevD.49.1944
  [hep-ph/9307203].

\bibitem{Wenzel:2007uh} 
  S.~Wenzel, E.~Bittner, W.~Janke and A.~M.~J.~Schakel,
  ``Percolation of Vortices in the 3D Abelian Lattice Higgs Model,''
  Nucl.\ Phys.\ B {\bf 793}, 344 (2008)
  doi:10.1016/j.nuclphysb.2007.10.024
  [arXiv:0708.0903 [hep-lat]].


\bibitem{Gleiser:1992ed} 
  M.~Gleiser and R.~O.~Ramos,
  ``Thermal fluctuations and validity of the one loop effective potential,''
  Phys.\ Lett.\ B {\bf 300}, 271 (1993)
  doi:10.1016/0370-2693(93)90365-O
  [hep-ph/9211219].

\bibitem{Ramos:1996at} 
  R.~O.~Ramos,
  ``Subcritical fluctuations at the electroweak phase transition,''
  Phys.\ Rev.\ D {\bf 54}, 4770 (1996)
  doi:10.1103/PhysRevD.54.4770
  [hep-ph/9607417].

\bibitem{Gleiser:2001gy} 
  M.~Gleiser, R.~Howell and R.~O.~Ramos,
  ``Dynamical precursor model for the onset of percolation,''
  Phys.\ Rev.\ E {\bf 65}, 036113 (2002)
  doi:10.1103/PhysRevE.65.036113
  [cond-mat/0106174].

\bibitem{Mocsy:2004ab} 
  A.~Mocsy, I.~N.~Mishustin and P.~J.~Ellis,
  ``Role of fluctuations in the linear sigma model with quarks,''
  Phys.\ Rev.\ C {\bf 70}, 015204 (2004)
  doi:10.1103/PhysRevC.70.015204
  [nucl-th/0402070].

\bibitem{Dziarmaga:1998js} 
  J.~Dziarmaga and M.~Sadzikowski,
  ``Anti-baryon density in the central rapidity region of a heavy ion collision,''
  Phys.\ Rev.\ Lett.\  {\bf 82}, 4192 (1999)
  doi:10.1103/PhysRevLett.82.4192
  [hep-ph/9809313].

\bibitem{Brandenberger:1998ew}
  R.~H.~Brandenberger and X.~m.~Zhang,
  ``Anomalous global strings and primordial magnetic fields,''
  Phys.\ Rev.\ D {\bf 59}, 081301  (1999) 
  doi:10.1103/PhysRevD.59.081301
  [hep-ph/9808306].
  
\bibitem{Nagasawa2}
M.~Nagasawa and R.~Brandenberger,
  ``Stabilization of the electroweak Z string in the early universe,''
  Phys.\ Rev.\ D {\bf 67}, 043504 (2003)
  doi:10.1103/PhysRevD.67.043504
  [hep-ph/0207246].
  
\end{thebibliography}
\end{document}